\documentclass[]{interact}

\usepackage{epstopdf}
\usepackage[caption=false]{subfig}

\usepackage{natbib}
\usepackage{amsmath}
\usepackage{graphicx,psfrag,epsf}
\usepackage{enumerate}
\usepackage{url} 
\usepackage{float}
\usepackage{amssymb}
\usepackage{xcolor}
\usepackage{caption}
\usepackage{tikz}
\usepackage{bm}
\usepackage{xcolor} 
\usepackage{hyperref}
\usepackage{geometry}
\definecolor{darkblue}{rgb}{0,0,0.5} 
\definecolor{darkred}{rgb}{0.5,0,0} 

\bibpunct[, ]{(}{)}{;}{a}{}{,}
\renewcommand\bibfont{\fontsize{10}{12}\selectfont}

\theoremstyle{plain}

\theoremstyle{definition}

\theoremstyle{remark}

\begin{document}

\articletype{}

\title{Hierarchical Count Echo State Network Models with Application to Graduate Student Enrollments}

\author{Qi Wang, Paul A. Parker, and Robert Lund \\
Department of Statistics \\
University of California, Santa Cruz}
\date{today}

\maketitle

\begin{abstract}
Poisson autoregressive count models have evolved into a time series staple for correlated count data. This paper proposes an alternative to Poisson autoregressions: count echo state networks. Echo state networks can be statistically analyzed in frequentist manners via optimizing penalized likelihoods, or in Bayesian manners via MCMC sampling. This paper develops Poisson echo state techniques for count data and applies them to a massive count data set containing the number of graduate students from 1,758 United States universities during the years 1972-2021 inclusive. Negative binomial models are also implemented to better handle overdispersion in the counts.  Performance of the proposed models are compared via their forecasting performance as judged by several methods. In the end, a hierarchical negative binomial based echo state network is judged as the superior model. 
\end{abstract}


\newpage
\section{Introduction}
\label{sec:intro}

Correlated count models are often used to describe positive integer-valued data, such as infectious disease counts \cite{agosto2020poisson}, bank failures \citep{schoenmaker1996contagion}, storm frequencies \citep{robbins2011changepoints}, hospital visits \citep{neelon2013spatial, matteson2011forecasting}, and crime incidents \citep{kim2021robust}. Standard Gaussian time series models may not describe discrete counts well, especially when the counts are small \citep{davishandbook}. As such, a discrete distribution is often adopted for the marginal distribution of the counts; Poisson, generalized Poisson, Conway-Maxwell Poisson, binomial, and negative binomial are common choices. The objective then becomes building correlated but non-Gaussian models for the counts. 

Classical approaches to the count problem include integer autoregression and general thinning approaches \citep{jin1991integer, jung2006binomial, weiss2008thinning, zhu2010negative, joe2016markov}, GLARMA techniques whereby model parameters evolve in a stationary but random fashion \citep{dunsmuir2015generalized}, Poisson autoregressive methods \citep{fokianos2009poisson}, and Gaussian transformations \citep{jia2023latent}. Recently, \cite{kong2024poisson} have written exclusively on count series having marginal Poisson distributions.  The above papers typically consider univariate series; literature considering multivariate counts is somewhat sparser, but includes\cite{ord1993time, heinen2007multivariate, brandt2012bayesian, serhiyenko2015approximate, karlis2016models, fokianos2021multivariate}.  Our setting, which contains over 1,700 count time series, lies in the high-dimensional realm where work is almost non-existent. The only papers we know studying high-dimensional counts are \cite{bradley2018computationally, pan2024modelling} and \cite{duker2024high}.

This paper studies modeling of high-dimensional count time series data with echo state networks (ESNs) (see \cite{mcdermott2017ensemble} for more on ESNs). The weights in our recurrent neural network are ``pre-generated" and fixed throughout training. Our count modeling tactics are similar to the GLARMA methods of \cite{dunsmuir2015generalized}. Penalized likelihood methods in frequentist settings and regression models with multivariate log-gamma priors in the Bayesian setting \citep{bradley2018computationally} are considered. For the negative binomial case, a P\'olya-gamma augmentation method is used to improve MCMC computational efficiency. To our knowledge, there is little literature on count ESNs except for \cite{schafer2020alternative}. Another innovation of this paper includes a count ESN model with a Bayesian hierarchical structure. Computationally efficient MCMC routines for Bayesian estimation are developed for both the Poisson and negative binomial models. 

Our interest in developing this methodology is for modeling of graduate student enrollment counts based on data from the Survey of Graduate Students and Postdoctorates in Science and Engineering, which is sponsored by the National Center for Science and Engineering Statistics, a federal statistical agency tasked with measurement and reporting of the U.S. science and engineering enterprise. The modeling of graduate student enrollment counts at individual schools within colleges across the United States fills a gap in the analysis of higher education data. Such analyses facilitate a deeper understanding of enrollment dynamics and provide valuable insights for evidence-based policy making, resource allocation, and institutional evaluation. This aligns with the objectives of official statistics, which strive to ensure accurate, transparent, and actionable information for various stakeholders. We also note that the methodology may be of broader interest to other statistical agencies that deal with count data over time. For example, the American Community Survey is used to understand counts such as the number of residents in a household (e.g., see \cite{parker2020conjugate}).

The rest of this paper proceeds as follows.  Section \ref{sec:data} describes the graduate student count data set that motivates this study.  Section \ref{sec:bkg} narrates the count time series and spatio-temporal modeling background needed to develop our model. Our ESN approach is developed in Section \ref{sec:methods}. Some competing models and scoring rules are given in Section \ref{sec:models}. Section \ref{sec:app} fits the ESN model to our graduate student counts, comparing to several other techniques.  Section \ref{sec:dis} summarizes our findings and proposes several directions for future work.

\section{Our Graduation Count Series}
\label{sec:data}
The data in this paper were extracted from the Survey of Graduate Students and Postdocs in Science and Engineering (GSS), an annual census of all academic institutions in the United States that grant research-based graduate degrees. This comprehensive data spans 1972-2021 inclusive and contains 1,758 schools. The GSS is a key source of information on demographics, study fields, support sources, and post-graduate plans of graduate students and postdoctoral researchers in selected fields of science, engineering, and health. The data are annual and allow us to examine time trends, patterns, and differences between institutions. 

The data collected in the GSS includes the number of graduate students and postdoctoral researchers by field of study, gender, citizenship status, and race/ethnicity. The survey also contains information on the primary sources of financial support for these individuals, such as federal agencies, universities, and private industry. Herein, we focus solely on the number of graduate students in each school. Note that schools (i.e., colleges) may be nested within institutions (i.e., universities). 

From the longitudinal nature of our data, it is possible to explore time changes in graduate student compositions. Model fitting methods could allow for trends or forecast counts in future years. The cross-sectional component of our data enables comparisons between institutions, providing insight into how different universities train and support our next generation of scientists and engineers.

\begin{figure}[htbp]
    \centering
    \includegraphics[width=1\linewidth]{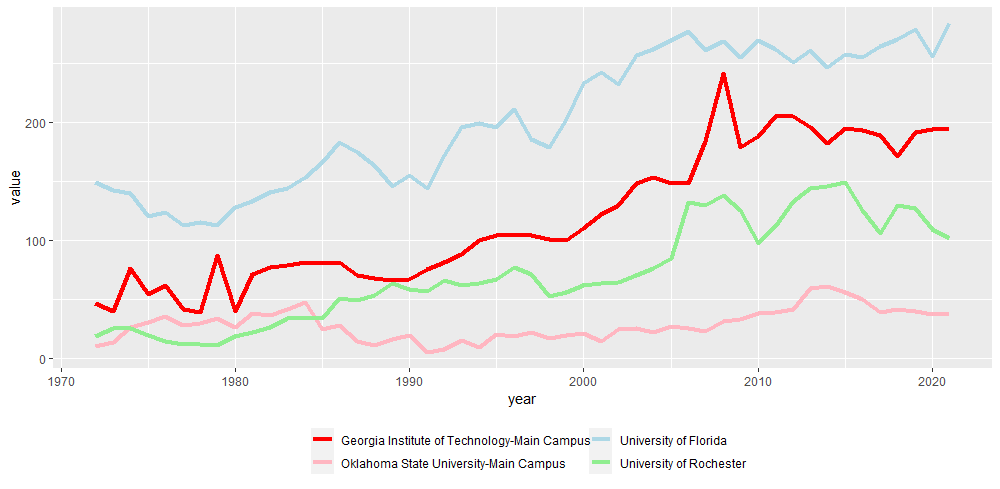}
\caption{Time series plots of four sample schools in our study.}
    \label{fig:tsexample}
\end{figure}

\begin{figure}[htbp]
    \centering
    \includegraphics[width=1\linewidth]{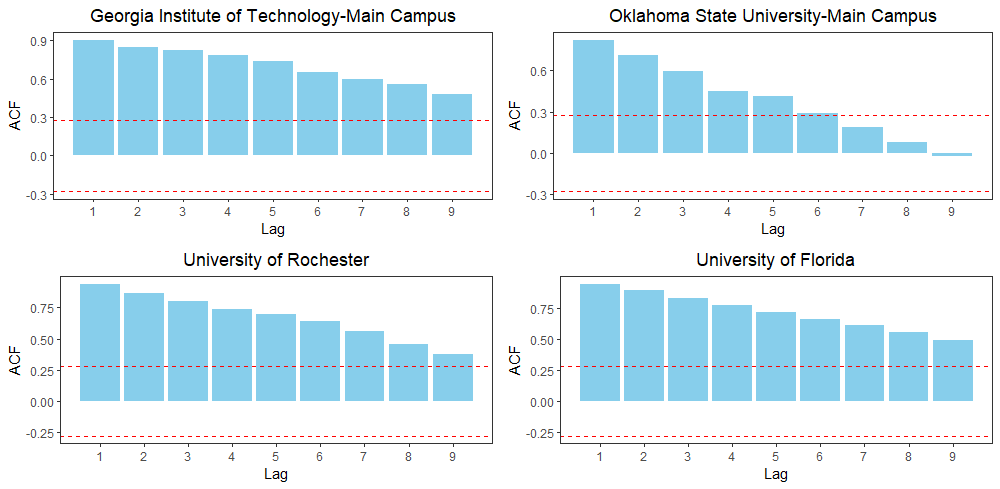}
    \caption{Sample autocorrelations for the four schools in Figure 1.  The dashed lines are pointwise 95\% confidence thresholds for a zero correlation.}
    \label{fig:acfexample}
\end{figure}

Figure \ref{fig:tsexample} plots time count trajectories for four schools in our study: The University of Florida (School of Chemistry), The University of Rochester (School of Electrical, Electronics, and Communications Engineering), The Georgia Institute of Technology (School of Chemical Engineering), and Oklahoma State University (School of Geological and Earth Sciences). The counts have varying scales; some schools have hundreds of students, while others report in teens. Sample autocorrelations for these four schools are shown in Figure \ref{fig:acfexample} and exhibit positive but non-negligible temporal associations.

\section{Background}
\label{sec:bkg}
This section reviews count time series and neural network methods.  Conjugacy approaches for our Bayesian models are also discussed. We begin with count time series.

\subsection{Count Time Series}

The classical way to model count time series is through integer autoregressions, which are based on  thinning operations. An integer autoregression of order one (INAR(1)) for the counts $\{ Y_t \}_{t=1}^T$ obeys
\[
Y_t = \delta \circ Y_{t-1} + \varepsilon_t,
\]
where $\{ \varepsilon_t \}$ is an IID positive integer-valued random sequence. The thinning operator $\circ$ operates on the integer-valued random varaible $N$ via  
\[
\delta \circ N = \sum_{i=1}^{N} B_i,
\]
where $\{ B_i \}$ are IID Bernoulli trials having success probability $\delta \in [0,1]$. One often selects $\{ \varepsilon_t \}$ to produce a specific count marginal distribution for $Y_t$. For example, selecting $\varepsilon_t$ to be Poisson with mean $\lambda(1-\delta)$ will produce a Poisson series with mean $\lambda$ for every $t$. Not every marginal distribution can be made; \cite{joe2016markov} discusses which distributions can be made. INAR models cannot have any negative autocorrelations, which does not appear to be detrimental here given the plots in Figure \ref{fig:acfexample}. Prominent INAR references include \cite{jin1991integer, jung2006binomial, weiss2008thinning}, and \cite{zhu2010negative}.

A more general count modeling approach was recently introduced in \cite{jia2023latent}. This approach transforms a standardized Gaussian series into the desired count series and can accommodate any marginal distribution.  Suppose that we seek a series having the marginal cumulative distribution function (CDF) $F_Y(\cdot)$. If $\{ Z_t \}$ is a standard Gaussian series with $E[ Z_t ] \equiv 0$, $\mbox{Var}(Z_t) \equiv 1$, and $\mbox{Corr}(Z_t, Z_{t+h})= \rho_Z(h)$, then set
\begin{equation}
\label{eqn:gt}
Y_t = F_Y^{-1}(\Phi(Z_t)),
\end{equation}
where $\Phi(\cdot)$ denotes the CDF of a standard normal distribution and $F^{-1}_Y(\cdot)$ is the quantile function
\[
F^{-1}_Y(u) = \inf \{ x: F_Y(x) \geq u \}.
\]
The probability integral transformation theorem shows that $\Phi(Z_t)$ has a uniform[0,1] distribution. A second application of the same result shows that $Y_t$ has the desired marginal distribution $F_Y(\cdot)$. 

The autocovariances of $\{ Y_t \}$ in (\ref{eqn:gt}) can be computed through Hermite expansions as in \cite{jia2023latent}. \cite{kong2024poisson} focuses exclusively on Poisson distributed series. 

State-space models are another popular count-modeling technique. A hierarchical model with some Poisson dynamics, for example, takes
\[
Y_t | \eta_t \sim \mbox{Poisson}(e^{\eta_t}),
\]
where $\{ \eta_t \}$ is a process to be clarified and a log link was used to keep the Poisson parameter positive at all times.  A Gaussian autoregressive structure is often placed on $\{ \eta_t \}$: 
\[
\eta_t = \phi_0 + \sum_{i=1}^p \phi_i \eta_{t-i} + \varepsilon_t.
\]
Here, $\{ \varepsilon_t \}$ is an IID Gaussian noise process with zero mean and $\phi_1, \ldots, \phi_p$ are the $p$ autoregressive coefficients that are assumed to produce a causal autoregression. While these and other models are discussed in \cite{fokianos2009poisson, davis2021count}, a useful tactic allows $\eta_t$ to depend on past counts. For example, in the first-order case, one could posit that 
\[
\eta_t = \phi_0 + a \eta_{t-1} +b Y_{t-1}.
\]
See also integer GARCH and Bayesian dynamic models for related work \citep{gamerman2015dynamic, holan2016hierarchical}.

\subsection{Neural Networks}

Neural networks are combinations of linear and nonlinear transformations, which often capably describe nonlinear features in sequential data. Neural networks that are based on the common recursions governing sequential data are called recurrent neural networks (RNNs) and were introduced in \citep{rumelhart1986learning}. Thereafter, \cite{hochreiter1997long} developed the so-called long short-term memory (LSTM) networks by adding additional structure that captures long-term dependence in the data. \cite{dey2017gate} simplified LSTMs while retaining similar model performance. 

Neural networks can be computationally intensive due to large parameter counts, especially in RNNs. ESNs, proposed by \cite{jaeger2004harnessing}, are variants of RNNs that providemore efficient model training. Hidden weights in the ESNs are randomly assigned as mixtures of zeros (the spike) and a symmetric distribution centered about zero (the slab). The only parameters requiring estimation reside in the output layer. The sparsity of the hidden layer and the reduction of unknown parameter numbers make ESNs computationally efficient. 

\cite{mcdermott2017ensemble} introduce a quadratic ESN (QESN) that models spatio-temporal data.  Uncertainty quantification is done via ensembling. A reduction layer is added to the QESN in \cite{mcdermott2019deep}, making it a deep-ensembled ESN; the model is implemented in a Bayesian framework. Furthering this work, \cite{schafer2020alternative} generalize ESNs to exponential families. \cite{wang2024echo} integrate graph convolutional neural networks and ESNs to capture areal-level spatial dependence. The details of our ESNs are developed below.

Recurrent neural networks (RNNs) are powerful modern techniques that can model time dependent sequential structures. The recurrent layers in RNNs view the inputs sequentially, retaining sequential dependence in the output.  

As an example, consider a single school from the GSS and let $\{ \bm{Y}_t \}_{t=1}^T$ denote the observations from the school. We have a length-$r$ covariate $\boldsymbol{x}_t$, which is school specific, at time $t$.  A RNN with $n_h$ hidden nodes can be built to fit this data. To begin, the length-$n_h$ vector $\boldsymbol{h}_1$ is initialized from the covariate $\boldsymbol{x}_1$.  Thereafter, for each $t = 2, 3, \ldots, T$, the hidden unit $\boldsymbol{h}_t$ in our recurrent layer is calculated via
\begin{equation}
\label{eqn:rnn}
\boldsymbol{h}_t = g(\boldsymbol{W}_h^\prime \boldsymbol{h}_{t-1} +
\boldsymbol{W}_{\boldsymbol{x}}^\prime \boldsymbol{x}_{t} +
\boldsymbol{c}). 
\end{equation}

\begin{figure}[H]
\begin{center}
\begin{tikzpicture}[scale = 0.6]
\draw[->, line width = 2pt] (-3,0) -- (-0.5,0);
\fill[darkblue] (0,0) circle [radius=0.5cm];
\node[text = white] at (0,0) {$\boldsymbol{h}_{t-1}$};
\fill[darkred] (1.5,2) rectangle (2.5,1);
\node[text = white] at (2,1.5) {$\boldsymbol{W}_h$};
\draw[->, line width = 2pt] (0.5,0) -- (4,0); 
\draw[->, line width = 2pt] (2,1) -- (2,0);

\fill[brown] (1,-2) circle [radius=0.5cm];
\node[text = white] at (1,-2) {$\boldsymbol{x}_t$};
\fill[darkred] (2.5,-3) rectangle (3.5,-4);
\node[text = white] at (3,-3.5) {$\boldsymbol{W}_x$};
\draw[->, line width = 2pt] (1.5,-2) -- (4,-2); 
\draw[->, line width = 2pt] (3,-3) -- (3,-2);

\draw[line width = 2pt] (4,1) rectangle (6, -3);
\fill[darkred] (4.5,3) rectangle (5.5,2);
\node[text = white] at (5,2.5){$\boldsymbol{c}$};
\draw[->, line width = 2pt] (5,2) -- (5,1);

\node at (5,-1){$\boldsymbol{g}(\cdot)$};
\draw[->, line width = 2pt] (6,-1) -- (7.5, -1);
\fill[darkblue] (8,-1) circle [radius=0.5cm];
\node[text = white] at (8,-1){$\boldsymbol{h}_t$};
\draw[->, line width = 2pt] (8.5,-1) -- (11,-1);
\draw (-2, 3.5) rectangle (10,-4.5);
\end{tikzpicture}
\end{center}
\caption{A recurrent layer example for sequentially calculating nodes}
\label{fig:rnn}
\end{figure}
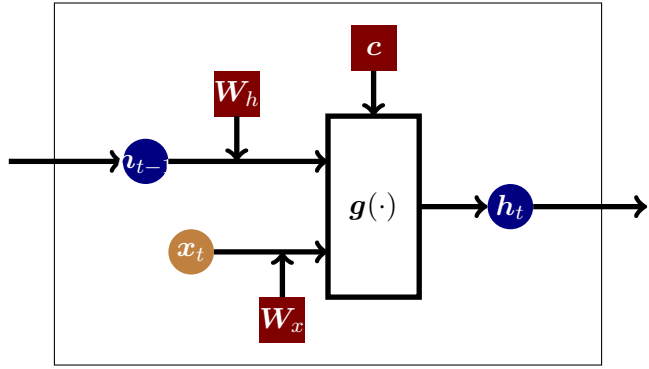

The $r \times n_h$ matrix $\boldsymbol{W}_{\boldsymbol{x}}$ affinely transforms our covariates at each time $t$. Recurrent layers use the $n_h \times n_h$ weight matrix $\boldsymbol{W}_{h}$ to affinely transform $\boldsymbol{h}_{t-1}$, the hidden layer output for the $(t-1)$th observation. The length-$n_h$ vector $\boldsymbol{c}$ is called an intercept or bias. The activation function $g(\cdot)$ is usually assumed bounded to avoid overflow issues. Common choices include the hyperbolic tangent or Sigmoid functions, which are (respectively)
\[
g(x) = \frac{e^{x} - e^{-x}}{e^x + e^{-x}},
\quad
g(x)=\frac{e^x}{1+e^x}.
\]
The calculation of $\boldsymbol{h}_t$ in a recurrent layer is graphically illustrated in Figure \ref{fig:rnn}. 

The parameters in the hidden layer of our RNN are $\boldsymbol{W}_h$, $\boldsymbol{W}_x$, and
$\boldsymbol{c}$. An output layer is added after the recurrent layer. The unknown parameters in both layers are estimated by numerically minimizing some loss function. Because both $\boldsymbol{x}_t$ and $\boldsymbol{h}_t$ are transformed and the effects in (\ref{eqn:rnn}) are summed, RNNs are very flexible models for temporal data. Next, we move to ESNs, which show how to take advantage of randomization to reduce the computational burden of RNNs.

An ESN's hidden layer weights are pre-generated and fixed throughout. The recurrent layer is intended to capture time dependence and the pre-generated weight matrix significantly reduces computations. \cite{mcdermott2017ensemble} show how to use an ESN to model time-dependent data. 

The observation for the $i$th school at time $t$, $Y_{i,t}$, has the conditional mean
\begin{gather}
\label{eqn:esnmodel}
E[ Y_{i,t} | Y_{i,1}, \ldots, Y_{i,t-1}] = \boldsymbol{h}_{i,t}^\prime \boldsymbol{\eta},
\end{gather}
where 1) the $n_h \times 1$ vector $\boldsymbol{h}_{i,t}$ denotes the output from a fixed-weight recurrent layer with $n_h$ hidden nodes and depends on previous counts for this school, and 2) $\boldsymbol{\eta}$ is a deterministic length $n_h$ vector of regression coefficients. 

The ESN models $\boldsymbol{h}_{i,t}$ recursively in time via 
\begin{gather}
    \begin{aligned}
    \boldsymbol{h}_{i,t} = 
     g \left(\frac{\nu}{|\lambda_W|}\boldsymbol{W}^\prime\boldsymbol{h}_{i,t-1} + 
    \boldsymbol{U}_Y^\prime\boldsymbol{Y}_{i,(t-p):(t-1)}+ \boldsymbol{U}_X^\prime\boldsymbol{x}_{i,t}\right),
    \end{aligned}
    \label{eqn:esnhidden}
\end{gather}
where $g(\cdot)$ is an activation function, $\boldsymbol{W}$ is an $n_h \times n_h$ weight matrix, and $\boldsymbol{U}_Y$ is a $p \times n_h$ matrix, where $p$ is the autoregressive order (larger $p$s are typically indicative of longer temporal memory). Similarly, $\boldsymbol{U}_X$ is an $r \times n_h$ matrix. The parameters $\boldsymbol{W}$, $\boldsymbol{U}_Y$, and $\boldsymbol{U}_X$ are randomly generated from some specified distribution, but thereafter are fixed and not estimated in the model fitting procedure. Our notation uses $\boldsymbol{Y}_{i,(t-p):(t-1)}=(Y_{i,t-p}, \ldots, Y_{i,t-1})^\prime$. We assume that $p$ is known; in fact, the sequel uses $p=1$ exclusively.

Elaborating further, the elements $W_{i,j}$, $(U_Y)_{i,j}$, and $(U_X)_{i,j}$, for $1 \leq i,j \leq n_h$, are independently generated from the following spike and slab distributions:
\begin{gather}
    \begin{aligned}
    \label{eqn:weightdistesn}
    (W)_{i,j} &= \gamma_{i,j}^W \mbox{Unif}(-a_w, a_w) + (1-\gamma_{i,j}^W) \delta_0, \\
    (U_Y)_{i,j} &= \gamma_{i,j}^{U_Y} \mbox{Unif}(-a_{u_Y}, a_{u_Y}) + (1-\gamma_{i,j}^{U_Y})\delta_0, \\
    (U_X)_{i,j} &= \gamma_{i,j}^{U_X} \mbox{Unif}(-a_{u_X}, a_{u_X}) + (1-\gamma_{i,j}^{U_X})\delta_0,
    \end{aligned}
\end{gather}
where 
\[
\gamma_{i,j}^W     \sim \mbox{Bern}(\pi_w),
\gamma_{i,j}^{U_Y} \sim \mbox{Bern}(\pi_{u_Y}),
\gamma_{i,j}^{U_X} \sim \mbox{Bern}(\pi_{u_X}).
\]
Here, $\delta_0$ denotes a unit point mass at zero. The elements $\gamma_{i,j}^W$ are IID, as are $\gamma_{i,j}^{U_Y}$ and $\gamma_{i,j}^{U_x}$. The hyperparameters $\pi_w$, $\pi_{u_Y}$, and $\pi_{u_X}$ are given and not estimated. The length $p$ vector $\boldsymbol{Y}_{i,(t-p):(t-1)}$ contains the past $p$ observations for the $i$th school; i.e., $( Y_{i,t-1}, \ Y_{i,t-2},..., Y_{i,t-p} )^\prime$. The scalar $\lambda_W$ is the largest eigenvalue of $\boldsymbol{W}$ and $\nu$ is a regularization parameter between 0 and 1, used to ensure that $\boldsymbol{h}_{i,t}$ does not explode/overflow. Again, $g(\cdot)$ is an activation function. This model contains hyperparameters that can be chosen through cross-validation; however, they are not estimated: these parameters are $a_.$, $\nu$, and $\pi_.$. After fixing these hyperparameters, the weight matrices $\boldsymbol{W}$, $\boldsymbol{U}_Y$, and $\boldsymbol{U}_X$ are generated. The only parameter estimated in (\ref{eqn:esnmodel}) is $\bm{\eta}$. This setup can be interpreted as first feeding information into the ESN and using the hidden layer output output as explanatory variables to fit a regression. \cite{mcdermott2017ensemble} use a ridge penalty to avoid overfitting when estimating $\boldsymbol{\eta}$. One may also add empirical orthogonal functions (EOFs) or other basis functions as covariates in (\ref{eqn:esnhidden}) to capture cross-sample dependence as in \cite{mcdermott2017ensemble}. In all future ESN model fits, the ranges $a_{\cdot}$ in (\ref{eqn:weightdistesn}) were 0.01, the dropout probabilities $\pi_{\cdot}$ were 0.1, the number of hidden nodes $n_h$ is 30, and $\nu$ is 0.9. Elaborating, these choices were made based on cross-validation performance. The candidates for $a_{\cdot}$ were $0.01$, $0.1$, and $1$; for $\pi_{\cdot}$, choices include $0.1$, $0.3$, and $0.5$; candidate values for $n_h$ are $30$, $50$, $100$, and $120$; finally, for $\nu$, we tried $0.5$, $0.7$, and $0.9$.

\subsection{Multivariate log-Gamma Priors}
This subsection introduces the multivariate log-Gamma (MLG) distribution in \cite{bradley2018computationally} and \cite{bradley2020bayesian}. This distribution's conjugacy properties are convenient for count modeling. To simulate from this distribution, an $m$-dimensional random vector $\boldsymbol{w} = (w_1, \ldots, w_m)^\prime$ is first generated with mutually independent components: $w_i \sim \mbox{LG}(\alpha_i,\kappa_i)$. Here, $\mbox{LG}(\alpha,\kappa)$ denotes the log-Gamma distribution, which is the logarithm of a Gamma draw with shape parameter $\alpha$ and scale parameter $\kappa$. Then set
\[
\boldsymbol{q} = \boldsymbol{\mu} + \boldsymbol{V} \boldsymbol{w},
\]
where $\boldsymbol{V} \in \mathcal{R}^m \times \mathcal{R}^m$ and $\boldsymbol{\mu}=(\mu_1, \ldots, \mu_m)^\prime$ are deterministic (hierarchical structures will be later placed on these parameters). We call $\boldsymbol{q}$ a multivariate log-gamma random vector and write $\boldsymbol{q} \sim \mbox{MLG}(\boldsymbol{\mu}, \boldsymbol{V}, \boldsymbol{\alpha}, \boldsymbol{\kappa})$. Here, $\boldsymbol{q}$ has the probability density 
\[
f(\boldsymbol{q}|\boldsymbol{\mu},\boldsymbol{V},\boldsymbol{\alpha},\boldsymbol{\kappa}) = 
\frac{1}{ \det(\boldsymbol{V}\boldsymbol{V}^\prime)^{1/2}}
\left( \prod_{i=1}^m 
\frac{\kappa_i^{\alpha_i}}{\Gamma(\alpha_i)}
\right) \exp[\boldsymbol{\alpha}^\prime\boldsymbol{V}^{-1}(\boldsymbol{q-\mu})-\boldsymbol{\kappa}^\prime \exp\{ \boldsymbol{V}^{-1}(\boldsymbol{q}-\boldsymbol{\mu})\}]
\]
over $\mathcal{R}^m$.

\cite{bradley2018computationally} show that a multivariate log-Gamma random variable with parameters $(\boldsymbol{c}, \alpha^{1/2}\boldsymbol{V}, \alpha \boldsymbol{1}, \alpha \boldsymbol{1})$ converges in distribution to a multivariate normal distribution with mean $\boldsymbol{c}$ and covariance matrix $\boldsymbol{V}\boldsymbol{V}^\prime$ as $\alpha \rightarrow \infty$, noting also that $\alpha=1,000$ is typically sufficiently large to approximate this convergence. Accordingly, we set $\alpha=1,000$ throughout.  

One can also partition a MLG iate $\boldsymbol{q}$ with parameters $(\boldsymbol{\mu}, \boldsymbol{V}, \boldsymbol{\alpha}, \boldsymbol{\kappa})$ into an $r$-dimensional vector $\boldsymbol{q}_1$ and an $n-r$-dimensional vector $\boldsymbol{q}_2$ via $\boldsymbol{q} = (\boldsymbol{q}_1^\prime, \boldsymbol{q}_2^\prime)^\prime$. Here, $\boldsymbol{V}^{-1}$ can also be partitioned into $[\boldsymbol{H} | \boldsymbol{B}]$, where $\boldsymbol{H}$ is an $n \times r$ dimensional matrix and $\boldsymbol{B}$ is an $n \times n-r$ matrix. 

Our notation uses $\boldsymbol{q}_1|\boldsymbol{q}_2$ as a conditional multivariate log-gamma distribution:
\[
\boldsymbol{q}_1|\boldsymbol{q}_2  \sim \mbox{cMLG}( \boldsymbol{H},\boldsymbol{\alpha},\boldsymbol{\kappa}^*),
\]
which can be shown to have the probability density
\[
f(\boldsymbol{q}_1|\boldsymbol{q}_2) =
M \exp\{\boldsymbol{\alpha}^\prime\boldsymbol{H}\boldsymbol{q}_2 - {\boldsymbol{\kappa}^*}^\prime \exp(\boldsymbol{H}\boldsymbol{q}_1\},
\]
where $\boldsymbol{\kappa}^* = \exp\{\boldsymbol{B}\boldsymbol{q}_2-\boldsymbol{V}^{-1}\boldsymbol{\mu}-\ln(\boldsymbol{\kappa})\}$ and $M$ is a normalizing constant. 

\cite{bradley2018computationally,bradley2020bayesian} describe an efficient data augmentation strategy to sample from this distribution, an important step when building Bayesian hierarchical models with a Poisson conditional distributions. This will be discussed in more detail later.

\section{Methods}
\label{sec:methods}

This section proposes several ESN models for count data having various structures.  We begin by describing a general count ESN model. For school $i$, we pose the model
\begin{equation}
\label{Count_ESN}
Y_{i,t} | \boldsymbol{\theta}_{i,t}
\stackrel{ind}{\sim} F_{\theta_{i,t}},
\end{equation}
where $F$ is some count marginal distribution that depends on the parameter $\theta_{i,t}$.  Depending on the choice of $F$, $\theta_{i,t}$ may be multivariate. For example, $\theta_{i,t}$ is univariate when $F$ is Poisson and bivariate when $F$ is negative binomial. Our models allow $\boldsymbol{\theta}_{i,t}$ to evolve recursively in time $t$ according to an ESN. One example has $\theta_{i,t} = \boldsymbol{h}_{i,t}^\prime \boldsymbol{\eta}$, where $h_{i,t}$ evolves in $t$ as in (\ref{eqn:rnn}).

\subsection{Poisson Echo State Networks}
\label{ss: poisson ESN} 

Suppose that $Y_{i,t} | \theta_{i,t}$ is Poisson with mean $\theta_{i,t}$.   Our setup uses the log link
\[
Y_{i,t} | \theta_{i,t} \sim \mbox{Poisson}(e^{\theta_{i,t}})
\]
to keep the Poisson parameter positive.  Here, $\theta_{i,t} = \boldsymbol{h}_{i,t}^\prime \boldsymbol{\eta}_i$ is used, where $\{ \boldsymbol{h}_{i,t} \}$ evolves in time via an ESN. A first-order autoregressive example employing the logarithm of past observations posits 
\begin{gather}
\boldsymbol{h}_{i,t} = 
g \left( \frac{\nu}{|\lambda_W|}\boldsymbol{W}^\prime\boldsymbol{h}_{i,t-1} 
+ \ln(Y_{i,t-1}+1) \boldsymbol{U}_Y^\prime  
+\boldsymbol{U}_X^\prime \boldsymbol{x}_{i,t}\right), \quad t = 2, \ldots, T \nonumber \\
\boldsymbol{h}_{i,1} = g(\boldsymbol{U}_X^\prime \boldsymbol{x}_{i,1}). \nonumber
\end{gather}
In the above, the activation function $g(\cdot)$ is applied coordinate-wise to the quantities in parentheses, and unity is added to $Y_{i,t-1}$ to avoid taking a logarithm of zero. Here, $g(\cdot)$ is the hyperbolic tangent function, chosen to avoid overflow and introduce nonlinearity. Also, $\boldsymbol{U}_X$, $\boldsymbol{U}_Y$, and $\boldsymbol{W}$ are randomly generated from (\ref{eqn:weightdistesn}) and fixed throughout. The elements of $\boldsymbol{W}$ and $\boldsymbol{U}$ are generated from the spike and slab prior distribution in (\ref{eqn:weightdistesn}).  The only unknown parameter for the $i$th school is $\boldsymbol{\eta}_i$. 

Assuming conditional independence of $Y_{i,t}|\theta_{i,t}$ in $t$, the likelihood for the $i$th school, denoted by $\mathcal{L}_i(\boldsymbol{\eta}_i)$, is
\[
\mathcal{L}_i (\boldsymbol{\eta}_i)= 
\prod_{t=1}^T \frac{\exp(-e^{\theta_{i,t}})e^{\theta_{i,t}{Y_{i,t}}}}{Y_{i,t}!}.
\]
The parameter $\boldsymbol{\eta}_i$ can be estimated by maximizing the LASSO-based penalized log likelihood $\mathcal{L}_i^*$:
\[
\mathcal{L}_i^*(\boldsymbol{\eta}_i)=\sum_{t=1}^T \left[Y_{i,t} \theta_{i,t} - 
e^{\theta_{i,t}} \right]- \tau \sum_{j=1}^{n_h}|\eta_{i,j}|,
\]
where $\tau$ is a penalty parameter typically chosen by cross-validation. The weights in the ESN need only be generated once to fit the model, which we call a ``Single Poisson ESN." Alternatively, the ESN parameters can be generated multiple times, and the model fitted to each realization to construct an ``Ensemble ESN." Ensembling enables uncertainty quantification. The regularization parameter $\tau$ is chosen as 1.0 via cross-validation among $0.5$,  $1$, and $1.5$. Note, similar to \cite{mcdermott2017ensemble}, a small grid of candidate values was used here and predictive performance was generally strong, however, in some cases it may be worthwhile to use a finer grid of candidate values to further optimize prediction accuracy.

This model can also be implemented using a Bayesian framework. Conjugacy is ensured by specifying the multivariate log-gamma (MLG) prior above for $\boldsymbol{\eta}_i$:
\[
\boldsymbol{\eta}_i \sim \mbox{MLG}(\boldsymbol{0}_{n_h},\alpha^{1/2}\sigma_\eta\boldsymbol{I}_{n_h}, \alpha \boldsymbol{1}_{n_h},\alpha\boldsymbol{1}_{n_h}),
\]
where the distributional notation follows \cite{bradley2018computationally}, $\boldsymbol{I}_{k}$ denotes the $k \times k$ identity matrix, $\boldsymbol{1}_k$ denotes the $k$-dimensional vector containing all unit entries, and $\boldsymbol{0}_k$ is the $k$-dimensional vector containing all zeros.  The variance parameter $\sigma_\eta$ is chosen to be 0.1 for shrinkage. Note that $\sigma_\eta$ can also be random, which is discussed later. In this case, the joint density has the form
\begin{eqnarray*}
    p(\boldsymbol{Y}_i,\boldsymbol{\eta}_i) & \propto & \prod_{t=1}^T \left[ \exp(\theta_{i,t}Y_{i,t}-
    \exp(\theta_{i,t}) \right] \times \\     
~~~& & \exp \left( \alpha\boldsymbol{1}_{n_h}^\prime \alpha^{-1/2}\frac{1}{\sigma_\eta}\boldsymbol{I}_{n_h}\boldsymbol{\eta}_i 
- 
\alpha\boldsymbol{1}_{n_h}^\prime \exp\left[\alpha^{-1/2}\frac{1}{\sigma_\eta}\boldsymbol{I}_{n_h}\boldsymbol{\eta}_i\right] \right). \\
\end{eqnarray*} 
Consequently, the posterior distribution $p(\boldsymbol{\eta}_i|\boldsymbol{Y}_i)$ has the conditional MLG form
\begin{eqnarray*}
p(\boldsymbol{\eta}_i|\boldsymbol{Y}_i) &\propto& 
\exp \bigg\{
\left(\sum_{t=1}^T Y_{i,t}\boldsymbol{h}_{i,t}^\prime+\alpha\boldsymbol{1}_{n_h}^\prime \alpha^{-1/2}\frac{1}{\sigma_\eta}\boldsymbol{I}_{n_h}\right)\boldsymbol{\eta}_i  -\\
~ &~&
\exp\left[\sum_{t=1}^T\boldsymbol{h}_{i,t}^\prime\boldsymbol{\eta}_i\right] -
\alpha\boldsymbol{1}_{n_h}^\prime \exp\left[\alpha^{-1/2}\frac{1}{\sigma_\eta}\boldsymbol{I}_{n_h}\boldsymbol{\eta}_i\right]
\bigg\}. \\
\end{eqnarray*}
For notation, define $\boldsymbol{H}_i=(\boldsymbol{h}_{i,1}^\prime, \ldots, \boldsymbol{h}_{i,T}^\prime)^\prime$ and 
\[
\boldsymbol{L}_i = 
\begin{bmatrix}
\boldsymbol{H}_i \\
\alpha^{-1/2}\frac{1}{\sigma_\eta}\boldsymbol{I}_{n_h} 
\end{bmatrix}, \quad 
\boldsymbol{\xi}_i = 
\begin{bmatrix}
\boldsymbol{Y}_i^\prime &
\alpha \boldsymbol{1}_{n_h}^\prime 
\end{bmatrix}^\prime, \quad
{\rm and}
\quad \boldsymbol{\psi}_i= 
\begin{bmatrix}
\boldsymbol{1}_T^\prime &
\alpha \boldsymbol{1}_{n_h}^\prime 
\end{bmatrix}^\prime.
\]
Under this formulation, the posterior distribution is
\[
\boldsymbol{\eta}_i | \boldsymbol{Y}_i \sim \mbox{cMLG}(\boldsymbol{L}_i, \boldsymbol{\xi}_i,\boldsymbol{\psi}_i).
\]

\cite{bradley2018computationally,bradley2020bayesian} develop a computationally efficient procedure to sample from a CMLG distribution via a ``data augmentation" approach. The full model is a latent conjugate multivariate process (LCM) model \citep{bradley2020bayesian}. In this case, the model can be partitioned into two stages:
\begin{itemize}
    \item Data stage:
    \[
    Y_{i,t}|\boldsymbol{\eta}_i,\boldsymbol{q}_i\stackrel{ind}{\sim}\mbox{Poisson}(e^{\boldsymbol{h}_{i,t}^\prime\boldsymbol{\eta}_i+ \boldsymbol{b}_i^\prime \boldsymbol{q}_i})
    \]
    \item Parameter stage:
        \begin{eqnarray*}
           & \boldsymbol{\eta}_i|\boldsymbol{V},\boldsymbol{\alpha}_\eta,\boldsymbol{\kappa}_\eta,\boldsymbol{q}_i \sim \mbox{MLG}(-\boldsymbol{V}\boldsymbol{B}\boldsymbol{q}_i,\boldsymbol{V}, \boldsymbol{\alpha}_\eta,\boldsymbol{\kappa}_\eta)\\ 
            &f(\boldsymbol{q}_i) \propto 1
        \end{eqnarray*}
where
    \begin{eqnarray*}
        &\boldsymbol{V} = \alpha^{1/2}\sigma_\eta\boldsymbol{I}_{n_h}, \\
        &\boldsymbol{\alpha}_\eta = \alpha \boldsymbol{1}_{n_h},
        \boldsymbol{\kappa}_\eta = \alpha \boldsymbol{1}_{n_h}.
    \end{eqnarray*}
Here, $\boldsymbol{b}_i$ is a pre-specified $T$-dimensional vector, and the $n_h \times T$ matrix $\boldsymbol{B}$ is also pre-specified. The $T$-dimensional random vector $\boldsymbol{q}_i$ has an improper flat prior. Conditional on $\boldsymbol{q}_i=\boldsymbol{0}_{T}$, the likelihood is proportional to the original model. This result yields the following sampling approach.
    \end{itemize}

Sampling $\boldsymbol{\Tilde{\eta}}$ from the posterior distribution:
\begin{itemize}

\item Sample $\Tilde{\boldsymbol{\eta}}_i$ from the $\mbox{MLG}(\boldsymbol{0}, \boldsymbol{I}_{n_h}, \boldsymbol{\xi}_i, \boldsymbol{\psi}_i)$ distribution.

\item Affinely transform $\Tilde{\boldsymbol{\eta}}_i$ via
\begin{equation}
\label{eqn:cmlg}
(\boldsymbol{L}_i^\prime \boldsymbol{L}_i)^{-1}\boldsymbol{L}_i^\prime\Tilde{\boldsymbol{\eta}}_i    
\end{equation}
as a posterior sample of $\boldsymbol{\eta}_i|\boldsymbol{Y}_i$.
\end{itemize}

We now develop a Bayesian hierarchical model that incorporates similarities among schools within the same geographic state, which are likely to be subject to the same regulations and hence behave similarly. Our model specification is
\begin{gather}
Y_{i,t} | \theta_{i,t} \stackrel{ind}{\sim} 
\mbox{Poisson} (e^{\theta_{i,t}}), \nonumber \\
\theta_{i,t} = \boldsymbol{h}_{i,t}^\prime\boldsymbol{\eta}_{s(i)} + \delta_{s(i)}, \nonumber 
\end{gather}
where the ``hidden state" $\boldsymbol{h}_{i,t}$ evolves via
\begin{gather}
\boldsymbol{h}_{i,t} = 
g \left( \frac{\nu}{|\lambda_W|}\boldsymbol{W}^\prime\boldsymbol{h}_{i,t-1} 
+ \ln(Y_{i,t-1}+1) \boldsymbol{U}_Y^\prime  
+\boldsymbol{U}_X^\prime \boldsymbol{x}_{i,t}\right), \quad t = 2, \ldots, T \nonumber \\
\boldsymbol{h}_{i,1} = g(\boldsymbol{U}_X^\prime \boldsymbol{x}_{i,1}) \nonumber
\end{gather}
Here, the subscript $s(i)$ is the state of the United States containing the $i$th school. A hierarchical structure is imposed on both $\boldsymbol{\eta}_{s(i)}$ and $\boldsymbol{\delta}_{s(i)}$ as follows:
\begin{gather}
\boldsymbol{\eta}_j \sim \mbox{MLG} (\boldsymbol{0}_{n_h}, \alpha^{1/2}\sigma_\eta\boldsymbol{I}_{n_h}, \alpha \boldsymbol{1}_{n_h}, \alpha\boldsymbol{1}_{n_h}), \; j=1,\ldots,n_s \nonumber \\ \boldsymbol{\delta} \sim \mbox{MLG}(\boldsymbol{0}_{n_s},\alpha^{1/2}\sigma_\delta \boldsymbol{I}_{n_s}, \alpha \boldsymbol{1}_{n_s},\alpha\boldsymbol{1}_{n_s}), \nonumber 
\end{gather}
where $n_s=49$ is the number of states with schools in the GSS data, and the hyperparameters $\sigma_\eta$ and $\sigma_\delta$ follow half-Cauchy priors with the fixed scale parameter $\upsilon$:
\[
\sigma_\eta \sim \mbox{Half-Cauchy}(0,\upsilon),  \quad
\sigma_\delta \sim \mbox{Half-Cauchy}(0,\upsilon).
\]
In this paper, we implement a vague prior distribution by choosing $\upsilon=100$. Let $\boldsymbol{S}_i^\prime$ denote a length $n_s$ vector containing only zeroes and ones, where the $s(i)$th element is unity (and all other entries are zero). Then $\theta_{i,t} = \Tilde{\boldsymbol{h}}_{i,t}^\prime \Tilde{\boldsymbol{\eta}}$, where 
\begin{equation}
\label{eqn:mergeh}
\Tilde{\boldsymbol{h}}_{i,t} := 
(\boldsymbol{0}^\prime_{(s(i)-1) \cdot n_h},\boldsymbol{h}_{i,t}^\prime, \boldsymbol{0}^\prime_{(n_s-s(i)-1) \cdot n_h},\boldsymbol{S}_i^\prime)
, \quad {\rm and}, \quad 
\Tilde{\boldsymbol{\eta}} = (\boldsymbol{\eta}_1^\prime, \ldots, \boldsymbol{\eta}_{n_s}^\prime, \boldsymbol{\delta}^\prime)^\prime. 
\end{equation}

The prior distribution of $\Tilde{\boldsymbol{\eta}}$ is
\[
\Tilde{\boldsymbol{\eta}} \sim \mbox{MLG}
\left(\boldsymbol{0}_{(n_h+1)\times n_s},
\begin{bmatrix}
\sigma_\eta \boldsymbol{I}_{(n_h \cdot n_s)} & \boldsymbol{0}_{(n_h \cdot n_s) \times n_s} \\
\boldsymbol{0}_{n_s\times (n_h \cdot n_s)} & \sigma_\delta \boldsymbol{I}_{n_s}
\end{bmatrix},
\alpha \boldsymbol{1}_{(n_h+1) \cdot n_s},
\alpha \boldsymbol{1}_{(n_h+1) \cdot n_s}
\right).
\]
The joint density $\pi(\boldsymbol{Y}, \Tilde{\boldsymbol{\eta}}, \sigma_\eta, \sigma_\delta )$ can be expressed as
\begin{align}
\pi(\boldsymbol{Y},\Tilde{\boldsymbol{\eta}},&\sigma_\eta,\sigma_\delta ) \propto \nonumber \\
&\prod_{i=1}^N\prod_{t=1}^T \exp\left(\Tilde{\boldsymbol{h}}_{i,t}^\prime\Tilde{\boldsymbol{\eta}}_i y_{i,t} - \exp \left(\Tilde{\boldsymbol{h}}_{i,t}^\prime\Tilde{\boldsymbol{\eta}}_i\right)\right)\times 
\frac{1_{\{\sigma_\delta >0\}}}{1+(\sigma_\delta /\upsilon)^2} \times \frac{1_{\{\sigma_\eta>0\}}}{1+(\sigma_\eta/\upsilon)^2} \nonumber \\
&\times \left|\begin{array}{cc}
    \sigma_\eta^{-1} \boldsymbol{I}_{(n_h \cdot n_s)} & \boldsymbol{0}_{(n_h \cdot  n_s)\times  n_s} \\
    \boldsymbol{0}_{ n_s\times (n_h \cdot n_s)} & \sigma_\delta ^{-1} \boldsymbol{I}_{ n_s}
\end{array}\right| \times \exp\left( 
\alpha\boldsymbol{1}^\prime_{(n_h+1) \cdot  n_s}\cdot \alpha^{-1/2}
\begin{bmatrix}
    \sigma_\eta^{-1} \boldsymbol{I}_{(n_h\cdot n_s)} & \boldsymbol{0}_{(n_h \cdot  n_s)\times  n_s} \\
    \boldsymbol{0}_{ n_s\times (n_h\cdot n_s)} & \sigma_\delta^{-1} \boldsymbol{I}_{ n_s}
\end{bmatrix}\Tilde{\boldsymbol{\eta}} \right)\nonumber \\
&\times \exp\left( -\alpha\boldsymbol{1}_{(n_h+1)\cdot n_s}^\prime \exp\left(
\alpha^{-1/2} \begin{bmatrix}
    \sigma_\eta^{-1} \boldsymbol{I}_{(n_h\cdot n_s)} & \boldsymbol{0}_{(n_h \cdot  n_s)\times  n_s} \\
    \boldsymbol{0}_{ n_s\times (n_h\cdot n_s)} & \sigma_\delta^{-1} \boldsymbol{I}_{ n_s}
\end{bmatrix}\Tilde{\boldsymbol{\eta}} 
\right) \right). \nonumber
\end{align}
With the definitions
\begin{eqnarray*}
    &\boldsymbol{H} = (\Tilde{\boldsymbol{h}}_{1,1}^\prime, \ldots, \Tilde{\boldsymbol{h}}_{1,T}^\prime,\ldots, \Tilde{\boldsymbol{h}}_{S,1}^\prime, \ldots, \Tilde{\boldsymbol{h}}_{S,T}^\prime)^\prime ,\nonumber \\
    &\boldsymbol{Y} = (y_{1,1},\ldots,y_{1,T},\ldots, y_{S,1},y_{S,T})^\prime,
\end{eqnarray*}
the full conditional distribution of $\Tilde{\boldsymbol{\eta}}$ is still a conditional MLG:
\[
\Tilde{\boldsymbol{\eta}}|\cdot \sim \rm{cMLG}(\boldsymbol{L}, \boldsymbol{\xi}, \boldsymbol{\psi}),
\]
where
\begin{gather}
    \boldsymbol{L} = \begin{bmatrix}
        \boldsymbol{H} \\
        \alpha^{-1/2}\begin{bmatrix}
    \sigma_\eta^{-1} \boldsymbol{I}_{(n_h \cdot n_s)} & \boldsymbol{0}_{(n_h \cdot  n_s)\times  n_s} \\
    \boldsymbol{0}_{ n_s\times (n_h \cdot n_s)} & \sigma_\delta ^{-1} \boldsymbol{I}_{ n_s}
\end{bmatrix}
    \end{bmatrix}, \nonumber \\
    \boldsymbol{\xi} = (\boldsymbol{Y}^\prime,\alpha\boldsymbol{1}_{(n_h+1) \cdot n_s}^\prime),\nonumber \\
    \boldsymbol{\psi}_\eta = (\boldsymbol{1}_{NT}^\prime,\alpha\boldsymbol{1}_{(n_h+1) \cdot n_s}). \nonumber
\end{gather}
Note that $\boldsymbol{H}$ is an $(n_S \cdot n_T) \times ((n_h+1) \cdot n_s)$ matrix, therefore, the direct inversion in Equation \ref{eqn:cmlg} can be computationally intensive when sampling from the posterior distribution. One way to increase posterior sampling efficiency exploits the sparsity in $\boldsymbol{L}$. 

For $\sigma_\eta$ and $\sigma_\delta$, the full conditional distributions are
\begin{align}
p(\sigma_\eta|\cdot) &\propto  \nonumber \\
&\frac{1_{\{\sigma_\eta>0\}}}{1+(\sigma_\eta/\upsilon)^2} \times \sigma_\eta^{-(n_h\times n_s)}\times \exp(\alpha^{1/2}\sigma_\eta^{-1}\boldsymbol{1}_{n_h\times n_s}^\prime \Tilde{\boldsymbol{\eta}}_{[1:(n_h\times n_s)]}) \nonumber \\
&\times
\exp(-\alpha\boldsymbol{1}^\prime_{n_h \cdot n_s}\exp(\alpha^{-1/2}\sigma_\eta^{-1}\Tilde{\boldsymbol{\eta}}_{[1:(n_h\times n_s)]})) \nonumber
\end{align}
and
\[
p(\sigma_\delta|\cdot) \propto 
\frac{1_{\{\sigma_\delta>0\}}}{1+(\sigma_\delta/\upsilon)^2} \times \sigma_\delta^{-n_s}
\times 
\exp(\alpha^{1/2}\boldsymbol{1}_{ n_s}^\prime\sigma_\delta^{-1}\boldsymbol{I}_{n_s}\boldsymbol{\xi} - 
\alpha\boldsymbol{1}_{n_s}^\prime\exp(\alpha^{-1/2}\sigma_\delta^{-1}\boldsymbol{\xi})),
\]
where $\boldsymbol{\xi}$ contains the last $n_s$ elements in $\Tilde{\boldsymbol{\eta}}$.

\subsection{Negative Binomial Echo State Networks}
This subsection proposes a negative binomial model as a Poisson alternative, accounting for possible over-dispersion in conditional marginal distributions. Our setup ilizes the negative binomial distributional specification
\[
Y_{i,t} | r_{i},p_{i,t} \stackrel{ind}{\sim} \mbox{NB} (r_{i},p_{i,t}),
\]
with the probability mass function
\[
P(Y_{i,t}=y_{i,t})=\frac{\Gamma(y_{i,t}+r_{i})}{\Gamma(r_{i})\Gamma(y_{i,t}+1)}\left(p_{i,t}\right)^{y_{i,t}}\left(1-p_{i,t}\right)^{r_{i}}.
\]

We model $p_{i,t}$ via the logit transformation
\begin{gather}
 \nonumber \\
    \ln\left(\frac{p_{i,t}}{1-p_{i,t}}\right) = \boldsymbol{h}_{i,t}^\prime\boldsymbol{\eta}_{s(i)} + \boldsymbol{\delta}_{s(i)}
    , \nonumber 
    \end{gather}
where
\begin{gather}
\boldsymbol{h}_{i,t} = 
g \left( \frac{\nu}{|\lambda_W|}\boldsymbol{W}^\prime\boldsymbol{h}_{i,t-1} 
+ \ln(Y_{i,t-1}+1) \boldsymbol{U}_Y^\prime  
+\boldsymbol{U}_X^\prime \boldsymbol{x}_{i,t}\right), \quad t = 2, \ldots, T \nonumber \\
\boldsymbol{h}_{i,1} = g(\boldsymbol{U}_X^\prime \boldsymbol{x}_{i,1}) \nonumber
\end{gather}

For estimation, the likelihood function for this model is
\begin{align}
\label{eqn:nbllh}
    \mathcal{L}(\Tilde{\boldsymbol{Y}}_{i,t}|\Tilde{\boldsymbol{\eta}},r_{i})
    \propto
\frac{\Gamma(y_{i,t}+r_{i})}{\Gamma(r_{i})\Gamma(y_{i,t}+1)}\frac{\exp(\Tilde{\boldsymbol{h}}_{i,t}^\prime\Tilde{\boldsymbol{\eta}})^{r_{i}}}{\left(1+\exp(\Tilde{\boldsymbol{h}}_{i,t}^\prime\Tilde{\boldsymbol{\eta}})\right)^{y_{i,t}+r_{i}}},
\end{align}
where $\Tilde{\boldsymbol{h}}_{i,t}$ and $\Tilde{\boldsymbol{\eta}}$ are as in (\ref{eqn:mergeh}). In a Bayesian framework, P\'olya-Gamma data augmentation can be employed to aid in posterior sampling. 
By Equation (2) in \cite{polson2013bayesian}, the second fraction in (\ref{eqn:nbllh}) is
\[
2^{-b}e^{\varkappa_{i,t}\psi_{i,t}}\int_{0}^{\infty}e^{-\omega_{i,t}\psi_{i,t}^2/2}p(\omega_{i,t})d\omega_{i,t},
\]
where $b_{i,t}=Y_{i,t}+r_{i}$, $\varkappa_{i,t}= r_{i}-(y_{i,t}+r_{i})/2$, and $\omega$ follows a P\'olya-Gamma($b_{i,t},0$) distribution with $\psi=\Tilde{\boldsymbol{h}}_{i,t}^\prime\Tilde{\boldsymbol{\eta}}$. Therefore, the likelihood is
\[
\frac{\Gamma(y_{i,t}+r_{i})}{\Gamma(r_{i})\Gamma(y_{i,t}+1)}2^{-(y_{i,t}+r_{i})}\exp(\varkappa_{i,t}\Tilde{\boldsymbol{h}}_{i,t}^\prime\Tilde{\boldsymbol{\eta}})\int_{0}^\infty \exp({-\omega_{i,t}\psi_{i,t}^2/2})p(\omega_{i,t})d\omega_{i,t}.
\]

Given a prior $\pi(\Tilde{\boldsymbol{\eta}})$, the full conditional for $\Tilde{\boldsymbol{\eta}}$ can be calculated via
\begin{align}
p(\Tilde{\boldsymbol{\eta}}|\cdot) 
&\propto
\pi(\Tilde{\boldsymbol{\eta}}) \exp\left\{-\frac{1}{2}(\boldsymbol{H}\Tilde{\boldsymbol{\eta}}-\boldsymbol{\zeta})^\prime\Omega(\boldsymbol{H}\Tilde{\boldsymbol{\eta}}-\boldsymbol{\zeta}) 
\right\}, \nonumber
\end{align}
where $\boldsymbol{H}$ is the combined $\Tilde{\boldsymbol{h}}_{i,t}$ by row, $\boldsymbol{\Omega}$ is a diagonal matrix with diagonal elements $(\omega_{1,1}, \ldots, \omega_{1,T}; \ldots ; \omega_{N,1}, \ldots, \omega_{N,T})$, and
\[
\boldsymbol{\zeta} = \left(\frac{\varkappa_{1,1}}{\omega_{1,1}},\ldots,\frac{\varkappa_{N,T}}{\omega_{N,T}}\right).
\]
If a multivariate normal prior with mean $\boldsymbol{\mu}_\eta$ and covariance matrix $\boldsymbol{\Sigma}_{\eta}$ is assumed for $\Tilde{\boldsymbol{\eta}}$, then
\[
\Tilde{\boldsymbol{\eta}}|\cdot \sim \mbox{MVN}(\boldsymbol{\mu}_\eta^*,\boldsymbol{\Sigma}_\eta^*),
\] 
where
\begin{gather}
\boldsymbol{\Sigma}_\eta^* = (\boldsymbol{H}^\prime\boldsymbol{\Omega}\boldsymbol{H}+\boldsymbol{\Sigma}_\eta^{-1})^{-1}, \nonumber \\
\boldsymbol{\mu}_\eta^* = \boldsymbol{\Sigma}_\eta^*(\boldsymbol{H}^\prime\boldsymbol{\varkappa}+\boldsymbol{\Sigma}_\eta^{-1}\boldsymbol{\mu}_\eta). \nonumber
\end{gather}
On the other hand,
\[
\omega_{i,t}|\cdot \sim \mbox{PG}(b_{i,t},\Tilde{\boldsymbol{h}}_{i,t}^\prime\Tilde{\boldsymbol{\eta}}),
\]
where PG denotes a P\'olya-Gamma distribution, and $\boldsymbol{\varkappa}$ is a vector of stacked $\varkappa_{i,t}$. A zero vector is chosen for $\boldsymbol{\mu}_\eta$ to shrink the parameters towards zero, and $\boldsymbol{\Sigma}_\eta$ is the diagonal matrix 
\[
\begin{bmatrix}
    \sigma_\eta^{2} \boldsymbol{I}_{(n_h \cdot n_s)} & \boldsymbol{0}_{(n_h \cdot  n_s) \times n_s} \\
    \boldsymbol{0}_{ n_s \times (n_h \cdot n_s)} & \sigma_\delta^{2} \boldsymbol{I}_{ n_s}
\end{bmatrix}.
\]
The prior distributions for $\sigma_\eta^2$ and $\sigma_\delta^2$ are inverse Gammas using the rate parameterization: $\sigma_\eta^2 \sim \mbox{IG}(\alpha_\eta,\beta_\eta)$ and $\sigma_\delta^2 \sim \mbox{IG}(\alpha_\delta,\beta_\delta)$. In this paper, $\alpha_{\cdot}=0.001$ and $\beta_{\cdot}=0.001$ are chosen to yield a vague prior. Finally, the inverse of the dispersion parameter $r_{i}$ follows the half Cauchy distribution
\[
\frac{1}{r_{i}} \sim \mbox{Half-Cauchy}(0,1).
\]

The joint distribution now follows as 
\begin{align}
&p(\boldsymbol{Y},\Tilde{\boldsymbol{\eta}},\boldsymbol{\omega},\sigma_\eta,\sigma_\delta,r_{i}) \propto  \nonumber \\ 
&\prod_{i=1}^N\prod_{t=1}^T
\left\{\frac{\Gamma(y_{i,t}+r_{i})}{\Gamma(r_{i})\Gamma(y_{i,t}+1)}2^{-(y_{i,t}+r_{i})}\right\} 
\exp\left\{-\frac{1}{2}(\boldsymbol{H}\Tilde{\boldsymbol{\eta}}-\boldsymbol{\zeta})^\prime\Omega(\boldsymbol{H}\Tilde{\boldsymbol{\eta}}-\boldsymbol{\zeta}) 
\right\} \times \nonumber \\
&|\boldsymbol{\Sigma}_\eta|^{-1/2}\exp\left(-\frac{1}{2}\Tilde{\boldsymbol{\eta}}^\prime\boldsymbol{\Sigma}_\eta^{-1}\Tilde{\boldsymbol{\eta}}\right) \times \prod_{i=1}^N\prod_{t=1}^T \mbox{PG}(\omega_{i,t}|b_{i,t},0)  \times \nonumber \\
&\mbox{IG}(\sigma_{\eta}|\alpha_\eta,\beta_\eta) \times \mbox{IG}(\sigma_{\delta}|\alpha_\delta,\beta_\delta)\times \mbox{Half-Cauchy}(1/r_{i}|0,1). \nonumber
\end{align}

\section{Model Comparisons and Scoring Procedures}
\label{sec:models}

This section discusses how we evaluate model fits (scoring criteria). One of the most commonly used scoring criteria involves the one-step-ahead mean squared prediction errors (MSPE), defined at time $t$ by 
\[
\mbox{MSPE}_t = \frac{1}{n_S}\sum_{i=1}^{n_S} 
(\hat{Y}_{i,t} - Y_{i,t})^2,
\]
where $\hat{Y}_{i,t}$ denotes the one-step-ahead prediction of the number of graduate students in school $i$ at time $t$. As discussed above, the type of one-step-ahead prediction used depends on the model type fitted. In a few cases, schools experience a large shift in their year-to-year graduate student counts. The MSPE is not robust to such outliers. To address this limitation, the mean squared logarithmic prediction error (MSLPE), defined as
\[
\mbox{MSLPE}_t = \frac{1}{N} \sum_{i=1}^{N} (\ln(\hat{Y}_{i,t}+1)- \ln(Y_{i,t}+1))^2,
\]
can be used.  Here, unity is added to observed and predicted counts to avoid taking a logarithm of zero. Smaller MSPEs and MSLPEs indicate better-fitting models. 

A measurement of the quality of the uncertainty quantification is the interval score (IS), defined as
\[
\mbox{IS}_t(\alpha) = \frac{1}{N}\sum_{i=1}^{N}
\left\{
(u_{i,t}-l_{i,t}) + \frac{2}{\alpha}(l_{i,t}- Y_{i,t})I_{[Y_{i,t} < l_{i,t}]} +
\frac{2}{\alpha}(Y_{i,t}-u_{i,t})I_{[Y_{i,t} > l_{i,t}]}
\right\},
\]
where $l_{i,t}$ and $u_{i,t}$ are the lower and upper bounds of the $1-\alpha$ prediction interval for school $i$ at time $t$, respectively.  Here, the 95\% prediction interval $\alpha=0.05$ is used. A lower interval score indicates a better fitting model.

Another uncertainty metric is the interval coverage rate (ICR), which measures the proportion of observations that fall in the $\alpha \times 100\%$ prediction interval:
\[
\mbox{ICR}(\alpha) = \frac{1}{N} \sum_{i=1}^N I_{[l_{i,t}<Y_{i,t}<u_{i,t}]}.
\]
ICR values close to $\alpha$ are indicative of an accurate model.

\section{Model Performance on the GSS Data}
\label{sec:app}

This section fits the proposed models to the GSS data and compares their performance. As a baseline, an intercept-only model for each school, which describes the scenario where the counts are modeled only through mean effects only, is fitted. Next, a Poisson INGARCH model \citep{ferland2006integer, fokianos2009poisson} is fitted via the \texttt{R} package \texttt{tscount} \citep{liboschik2017tscount}. The remaining fitted models are variants of ESNs. First, a Single Poisson ESN is fitted. This fit provides a point estimate, but no uncertainty quantification. Thereafter, an Ensemble Poisson ESN is fitted, which allows for both a point estimate and uncertainty quantification. A Bayesian Poisson ESN is also fitted. Finally, Bayesian hierarchical ESNs, which take into account dependence between schools residing within the same US state, are considered for both Poisson and negative binomial marginal distributions. 

The regularization parameter $\tau$ in frequentist ESNs is chosen as unity. For the Bayesian Poisson ESN, no burn-in or thinning was used due to the conjugacy. For the Bayesian Hierarchical NB ESN, we burn in 1,000 samples and then use/save every other sample in the generated Markov Chain. When sampling $r_i$, we again use Metropolis-Hastings. The proposal distribution is chosen as uniform having a mean of the current value and range $2(\min\{10,r_i\})$. For the Bayesian Hierarchical Poisson ESN, the first 500 samples are burned in, and every other sample is again saved. Since the posterior distribution of $\sigma_\eta$ and $\sigma_\delta$ are not in closed form, a Metropolis-Hastings step is needed. Due to the positivity of $\sigma_\eta$ and $\sigma_\delta$, the proposal distribution is taken as uniform centered at the current value and having range $2 \min \{0.5, \sigma_\delta \}$ and $2 \min \{0.5, \sigma_\eta \}$, respectively. Each Bayesian method generates 1,000 prediction samples.

One-step-ahead predictions for the number of graduate students from 2017 to 2021 were made. These predictions use all data up to the previous year analyzed. Specifically, data from 1972-2016 is used to predict the 2017 counts; data from 1972-2020 are used to predict the 2021 counts.

The GSS data are generally overdispersed. To see this, the mean of the sample mean counts of all 1,758 schools is 63.48 and the mean of the sample variances (a denominator of $n-1$ is used) is 1320.28, significantly more than the mean. Hence, Poisson marginal distributions, which have a unit dispersion, will be inadequate.  Because of this, a hierarchical negative binomial setup, which permits more overdispersion than a hierarchical Poisson setup, is considered. 

\subsection{One-step-ahead Predictions}

MSPE and MSLPE scores for various model fits are shown in Table \ref{tab:mspe} and \ref{tab:mslpe}. For MSPE scores, the Bayesian Hierarchical NB ESN performs best, followed by the Bayesian Hierarchical Poisson ESN, which also has a relatively small MSE. The INGARCH(1,1) model and the frequentist ESNs, including the Single ESN and Ensemble ESN, exhibit similar performance.

For MSLPE scores, the Bayesian Hierarchical NB ESN again has the best performance. Note the difference between the frequentist ESNs and the INGARCH scores: on a logarithmic scale, frequentist ESNs are outperformed by INGARCH models.   

\begin{table}[]
\centering
\begin{tabular}{l|c|c|c|c|c|c}
\hline
  & 2017 & 2018 & 2019 & 2020 & 2021 & 5 Year Average\\
\hline
Intercept & 9537 & 9233 & 9077 & 6282 & 7716 & 8369\\
\hline
Poisson INGARCH(1,1) & 1129 & 612 & 512 & 960 & 587 & 760\\
\hline
Single Poisson ESN & 942 & 572 & 487 & 1412 & 411 & 765\\
\hline
Ensemble Poisson ESN & 909 & 515 & 481 & 1420 & 422 & 749\\
\hline
Bayesian Poisson ESN & 1146 & 656 & 594 & 984 & 521 & 780\\
\hline
Bayesian Hierarchical NB ESN & 1011 & 369 & 375 & 935 & 415 & 621\\
\hline
Bayesian Hierarchical Poisson ESN & 924 & 333 & 288 & 1348 & 422 & 663\\
\hline
\end{tabular}
\caption{MSPEs for All Fitted Models.}
\label{tab:mspe}
\end{table}

\begin{table}[]
    \centering
\begin{tabular}{l|c|c|c|c|c|c}
\hline
  & 2017 & 2018 & 2019 & 2020 & 2021 & 5 Year Average\\
\hline
Intercept & 1.046 & 1.040 & 1.088 & 1.133 & 1.199 & 1.101\\
\hline
Poisson INGARCH(1,1) & 0.215 & 0.109 & 0.110 & 0.193 & 0.144 & 0.154\\
\hline
Single Poisson ESN & 0.242 & 0.168 & 0.159 & 0.224 & 0.188 & 0.196\\
\hline
Ensemble Poisson ESN & 0.243 & 0.156 & 0.161 & 0.239 & 0.190 & 0.198\\
\hline
Bayesian Poisson ESN & 0.221 & 0.123 & 0.130 & 0.199 & 0.136 & 0.162\\
\hline
Bayesian Hierarchical NB ESN & 0.205 & 0.086 & 0.098 & 0.177 & 0.121 & 0.137\\
\hline
Bayesian Hierarchical Poisson ESN& 0.213 & 0.085 & 0.097 & 0.182 & 0.141 & 0.144\\
\hline
\end{tabular}
\caption{MSLPEs for all fitted models.}
    \label{tab:mslpe}
\end{table}

For model fits having uncertainty quantification, Table \ref{tab:is} and \ref{tab:icr} present IS(0.05) and ICR(0.95) results. The INGARCH(1,1) model achieves the best five-year average IS(0.05). However, the Bayesian Hierarchical NB ESN model has an overall ICR closer to 0.95, suggesting that the Bayesian Hierarchical NB and the Bayesian Hierarchical Poisson ESNs intervals are wider. This could be due to the ESN's sensitivity to changepoint-type shifts in some of the series.  Although the prediction intervals are capable of covering such shifts, this comes at the cost of widening the intervals in other regions.

\begin{table}[]
    \centering
\begin{tabular}{l|c|c|c|c|c|c}
\hline
  & 2017 & 2018 & 2019 & 2020 & 2021 & 5 Year Average\\
\hline
Poisson INGARCH(1,1) & 309 & 165 & 148 & 237 & 176 & 207\\
\hline
Ensemble Poisson ESN & 507 & 388 & 358 & 491 & 348 & 418\\
\hline
Bayesian Poisson ESN & 550 & 409 & 388 & 476 & 368 & 438\\
\hline
Bayesian Hierarchical NB ESN & 440 & 239 & 236 & 354 & 259 & 305\\
\hline
Bayesian Hierarchical Poisson ESN & 448 & 226 & 235 & 385 & 268 & 312\\
\hline
\end{tabular}
\caption{Interval scores (0.05) for models with uncertainty quantification.}
\label{tab:is}
\end{table}

\begin{table}[]
\centering
\begin{tabular}{l|c|c|c|c|c|c}
\hline
  & 2017 & 2018 & 2019 & 2020 & 2021 & 5 Year Average\\
\hline
Poisson INGARCH(1,1) & 0.720 & 0.811 & 0.795 & 0.769 & 0.766 & 0.772\\
\hline
Ensemble Poisson ESN & 0.678 & 0.703 & 0.725 & 0.697 & 0.724 & 0.705\\
\hline
Bayesian Poisson ESN & 0.685 & 0.736 & 0.734 & 0.720 & 0.749 & 0.725\\
\hline
Bayesian Hierarchical NB ESN & 0.927 & 0.976 & 0.972 & 0.952 & 0.962 & 0.958\\
\hline
Bayesian Hierarchical Poisson ESN & 0.768 & 0.873 & 0.866 & 0.827 & 0.845 & 0.836\\
\hline
\end{tabular}
\caption{Interval coverage rates (0.95) for models with uncertainty quantification.}
\label{tab:icr}
\end{table}

\subsection{In-sample Model Adequacy Checking}

\begin{figure}[htbp]
\centering
\includegraphics[width=1\linewidth]{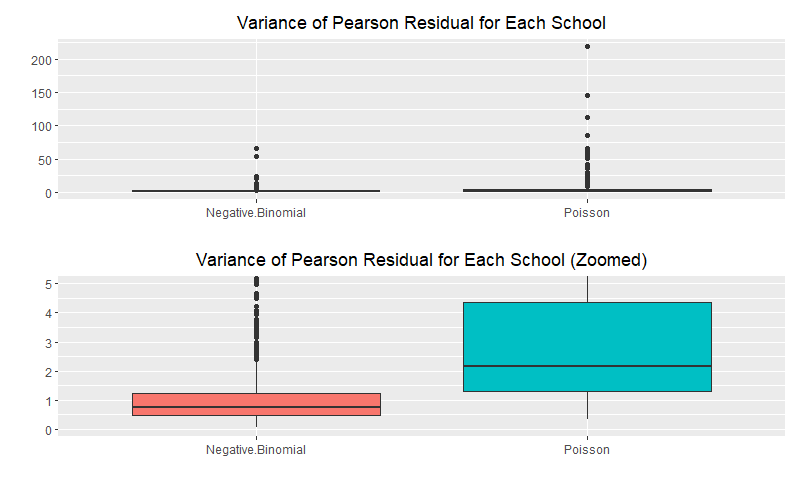}
\caption{Boxplot for the dispersions of all schools.}
\label{fig:disp_boxplt}
\end{figure}

To further assess the model fits, the conditional standardized Pearson residuals introduced by \cite{weiss2020checking} were computed. These residuals are
\[
R_{t}(\hat{\boldsymbol{\theta}}) := \frac
{Y_t - E(Y_t|Y_{t-1},Y_{t-2},\ldots,Y_1;\hat{\boldsymbol{\theta}})}
{\sqrt{\mbox{Var}(Y_t|Y_{t-1},Y_{t-2},\ldots,Y_1;\hat{\boldsymbol{\theta}})}}
\]
at time $t$.  If the fitted model is adequate, the computed $R_t$s should have a sample mean close to zero and a unit variance.  A boxplot of the sample variance of the residuals from each of the 1,728 schools is shown in Figure \ref{fig:disp_boxplt} for the Bayesian hierarchical NB ESN and Bayesian Hierarchical Poisson ESN fits.

\begin{figure}[htbp]
\centering
\includegraphics[width=1\linewidth]{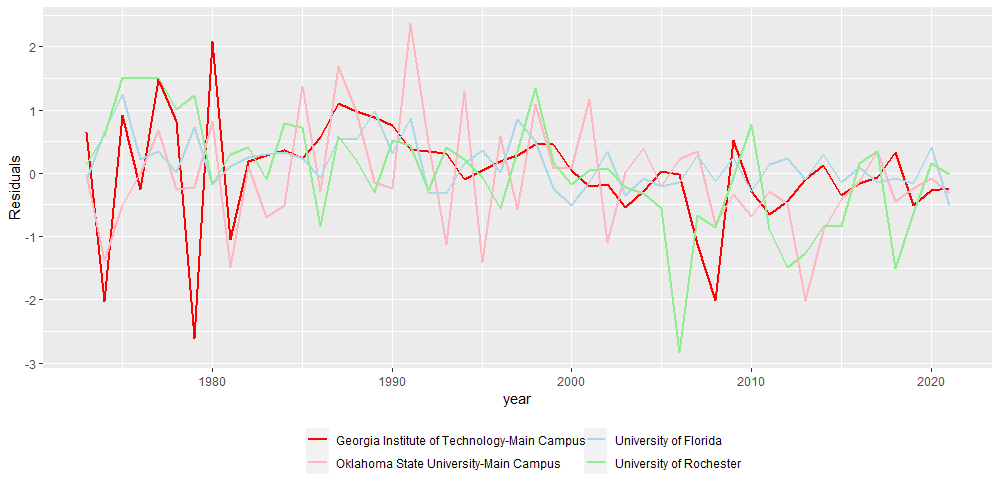}
\caption{Model residuals of the four schools in Figure \ref{fig:tsexample} from 1973-2021. The initial year 1972 is not considered due to edge effects.}
\label{fig:pred_res}
\end{figure}

\begin{figure}[htbp]
\centering
\includegraphics[width=1\linewidth]{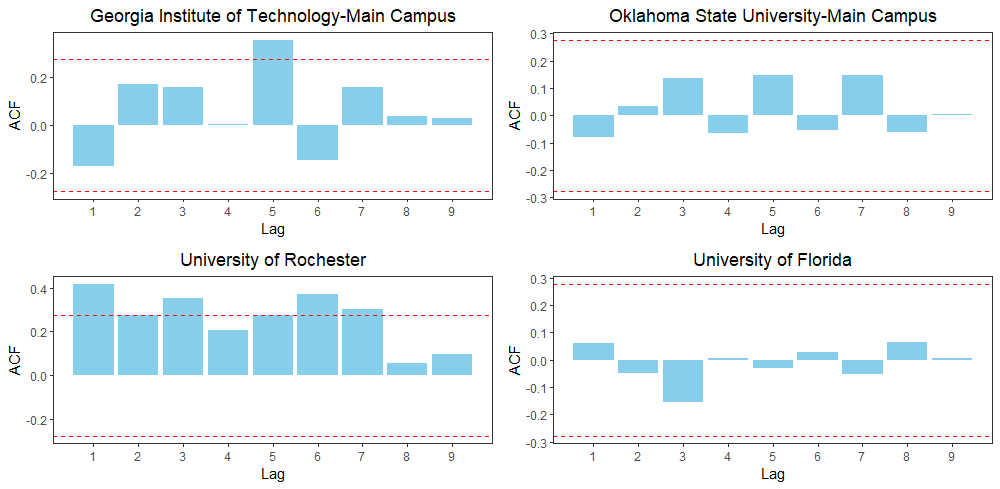}
\caption{Residual autocorrelations. The dashed lines are pointwise 95\% confidence thresholds for a zero autocorrelation.}
\label{fig:res_acf}
\end{figure}

The figure shows that the Bayesian Hierarchical NB ESN has a sample variance that is close to unity for most schools, and significantly outperforms the Poisson model. These residuals are further scrutinized in Figures \ref{fig:pred_res} and \ref{fig:res_acf}, which show time series plots and sample autocorrelations of the four schools in Figure \ref{fig:tsexample}. The residual series exhibit minimal autocorrelation overall, suggesting that the ESNs effectively capture temporal autocorrelations. However, for the University of Rochester, a few autocorrelations appear significantly different from zero. Given the number of schools and lags considered, this anomaly is not particularly concerning.

\section{Discussion}
\label{sec:dis}

This work developed various extensions of ESNs to model graduate student enrollment counts.  Parameters were estimated in both frequentist and Bayesian frameworks. Hierarchical structures were introduced to account for correlations from schools within the same geographic state of the US. One-step-ahead prediction residuals were assessed from 2017 to 2021 using various scoring metrics. Our enrollment count series were best modeled by a Bayesian hierarchical model with negative binomial dynamics. 

Future work aims to explore methods for capturing spatial effects in ways not considered here. \cite{wang2024spatial} propose using basis functions in a convolutional neural network to incorporate ``spatial effects" at different scales. One potential approach is to apply a spatial deep convolutional neural network (SDCNN) in an extreme learning machine framework \citep{huang2006extreme}. Here, the outputs from the SDCNN can be used as covariates. By combining SDCNN with ESNs, nonseparable and nonstationary spatial covariances (in either count or continuous cases) could possibly be handled. Another area where research is needed lies with marginal distribution types. Although only Poisson and negative binomial structures were considered here, research into an unspecified count family via nonparametric techniques could prove useful. 

\section*{Acknowledgement}
This research was partially supported by the U.S. National Science Foundation (NSF) under Grant NCSE-2215169. Robert Lund was partially supported by the NSF grant DMS-2113592. This article is released to inform interested parties of ongoing research and to encourage discussion. The views expressed on statistical issues are those of the authors and not those of the NSF.

\newpage
\bibliographystyle{chicago}

\end{document}